\documentclass[acmtog,authorversion,nonacm]{acmart}
\usepackage[utf8]{inputenc} 
\usepackage[T1]{fontenc}    

\usepackage{xcolor}
\usepackage{soul}
\usepackage{cleveref}
\usepackage{subcaption}
\usepackage{tikz}
\usepackage{enumitem}
\usepackage{algorithm}
\usepackage[noend]{algpseudocode}

\newtheorem{definition}{Definition}

\def\s{\sigma} 
\def\K{K} 
\def\L{L} 
\def\Face{\mathcal{F}} 
\def\M{\mathcal{M}} 

\begin{document}

\title{Codimensional MultiMeshing: Synchronizing the Evolution of Multiple Embedded Geometries}

\author{Michael Tao}
\email{michael.tao@nyu.edu}
\affiliation{%
    \institution{New York University}
    \country{United States}
}

\author{Jiacheng Dai}
\affiliation{%
  \institution{New York University}
  \country{USA}
}
\email{jd4705@nyu.edu}

\author{Denis Zorin}
\affiliation{%
\institution{New York University}
\country{United States}
}
\email{dzorin@cs.nyu.edu}

\author{Teseo Schneider}
\affiliation{%
  \institution{University of Victoria}
  \country{Canada}
}
\email{teseo@uvic.ca}

\author{Daniele Panozzo}
\affiliation{%
\institution{New York University}
\country{United States}
}
\email{panozzo@nyu.edu}

\renewcommand{\shortauthors}{Tao et al.}

\begin{abstract}
Complex geometric tasks such as geometric modeling, physical simulation, and texture parametrization often involve the embedding of many complex sub-domains with potentially different dimensions.
These tasks often require evolving the geometry and topology of the discretizations of these sub-domains,  and guaranteeing a \emph{consistent} overall embedding for the multiplicity of sub-domains is required to define boundary conditions.
We propose a data structure and algorithmic framework for hierarchically encoding a collection of meshes, enabling topological and geometric changes to be automatically propagated with coherent correspondences between them. We demonstrate the effectiveness of our approach in surface mesh decimation while preserving UV seams, periodic 2D/3D meshing, and extending the TetWild algorithm to ensure topology preservation of the embedded structures.
\end{abstract}

\begin{teaserfigure}\centering
     \includegraphics[width=\linewidth]{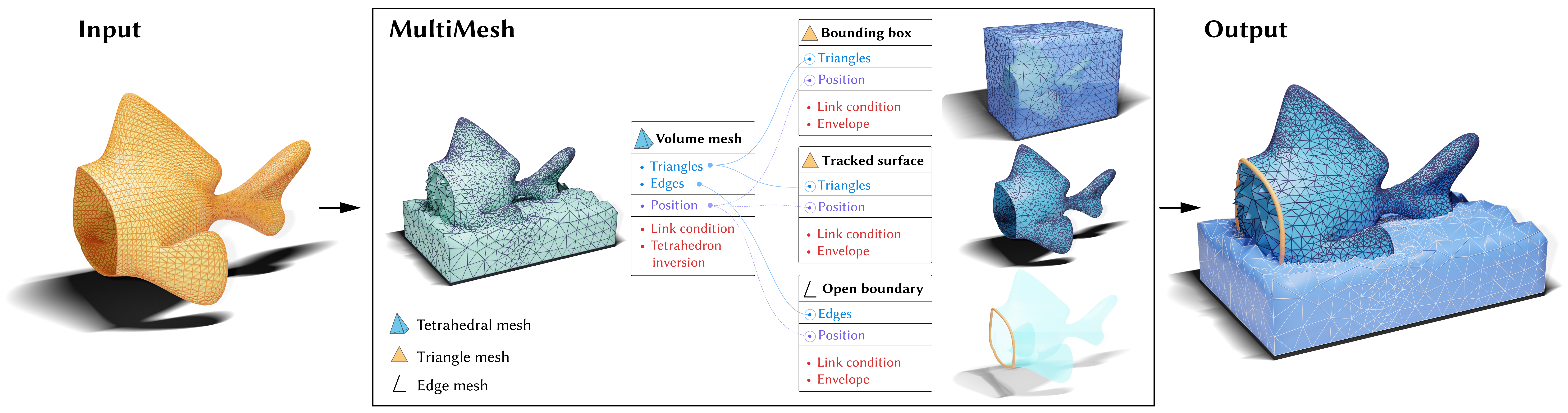}
      \caption{We use a multimesh composed of a volume (tetrahedra), a bounding box (triangles), an open surface (triangles), and the boundary of that open surface (edges) to run the TetWild optimization algorithm \cite{Hu:2018:TMW:3197517.3201353} on the tetrahedral mesh. We use the different sub-meshes to check for envelope containment instead of relying on tags. The multimesh data structure automatically keeps all meshes synchronized.}
  \label{fig:teaser}
\end{teaserfigure}

\maketitle

\section{Introduction}

A textured surface is often represented by two triangle meshes, one embedded in three dimensions for positions and one embedded in two dimensions for the texture.
Even though these two meshes often have the same number of triangles, those triangles are often connected differently, and these differences define the seams.
Similarly, a quadrilateral (hexahedral) mesh contains a lower dimensional mesh composed of vertices (singularities) and edges connecting them called a singularity graph. 
Finally, multi-material meshes, ubiquitous in bio-mechanical and fluid simulations, are often volumetric meshes with embedded surface meshes to represent the surfaces of separation between the materials. 

All these disparate examples have a common trait: they involve a collection of geometries of different dimensions related by a set of correspondences on how different elements correspond to one another.
In this article, we will discuss a systematic approach for maintaining these correspondences even as the geometries evolve and demonstrate its effectiveness on a set of graphics applications.

\paragraph{Tagging and Duplicated Attributes.}
In the literature, these cases are usually handled by maintaining an explicit representation of the highest dimensional mesh and defining a set of tags to encode the lower dimensional structures. 
In the case of multiple meshes of the same dimension, such as a surface mesh and its UV parametrization, the challenge is how the single mesh should store attributes representing different topologies. This is commonly done by redundantly storing the attributes, such as at the vertex corners of each triangle.
This practice leads to additional difficulty to keep the duplicated attributes synchronized. While this is sufficient for applications using static meshes, it becomes unwieldy as soon as mesh modifications are needed. For example, the optimization of the quality of a UV mesh while preserving the seam topology is a superficially simple problem, but a robust implementation of this is elusive in both commercial and open-source software due to its highly challenging implementation with per-corner attributes.

\paragraph{MultiMesh.} We propose a holistic modeling approach for all the cases above, by introducing the \emph{multimesh}. 
This data structure and algorithmic framework enables straightforward navigation between each mesh and robust editing of interconnected meshes, as if they were one mesh.
To the best of our knowledge, this problem, while very common in parametrization, meshing, and multi-material simulation, has never been directly investigated before.
After extensively considering alternatives (\Cref{sec:related}), we propose explicitly representing all meshes of all dimensions: each mesh has its independent navigation, attributes, and local editing operations. These meshes are interconnected by a tree of mappings which establishes a correspondence between simplices.

In \Cref{fig:teaser} we show an example of optimizing the quality of a tetrahedral mesh with two triangle meshes and an edge mesh embedded in it. The edge and triangle meshes provide envelope constraints (they cannot move too far from the input) and are kept up to date automatically during the optimization.

This is possible because the multimesh enables a dimension-agnostic navigation between meshes using a generalization of the darts/tuple formalism \cite{LIENHARDT1994}, perform local editing operations in any mesh, and automatically propagate the changes to the other  mesh in the multimesh. 
In \Cref{fig:operations_update}, we show that a split in an element is automatically propagated to all other linked simplices, eventually triggering additional topological changes to keep the involved meshes valid. An additional benefit of this approach is that algorithms designed for a mesh type (for example, isotropic remeshing on a surface mesh~\citep{Botsch2004}) can be directly applied when the mesh is embedded in another mesh without modification, enabling code reuse.

\paragraph{Applications}
We demonstrate the wide applicability of our approach to many classical graphics algorithms, which can be cleanly and compactly restated in our formulation. This includes the optimization of a surface mesh with seams, remeshing a periodic mesh, isotropic remeshing of a surface embedded in a tetrahedral mesh, and the re-implementation of the Tetwild mesh optimization algorithm.

\paragraph{Contributions}
Our contributions are:
\begin{itemize}
    \item A formal definition of a multimesh and the mechanisms for coherently maintaining it;
    \item An implementation of a multimesh and its operations;
    \item An extension of the declarative specification of \cite{toolkit} to a multimesh;
    \item An evaluation on a set of applications in graphics and scientific computing.
\end{itemize}

\section{Related Work}
\label{sec:related}

\subsection{Mesh Data Structures}

Data structures for encoding the connectivity of meshes have been researched for many years \cite{Requicha1980}: They differ in the types of simplices and the relationships between those simplices that are explicitly stored, and consequently the algorithms for navigating and editing the stored mesh differ as well.

The indexed data structure only stores attributes on the larger dimension simplices \cite{Botsch2010}, making it compact and a common choice for data exchange \cite{libigl}. 
Half-edges \cite{Mantyla1987,Baumgart1972,Guibas1985} or half-faces \cite{Dyedov2014} are more ideal for applications requiring heavy topological editing by encoding both sides of each edge or face with explicit references to adjacent simplices. Generalizations of this data structure are the generic cell tuple \cite{Brisson1989} and combinatorial maps \cite{Lienhardt1991,LIENHARDT1994,Feng2013}, that respectively represent the connectivity as a collection of ``tuples'' or ``darts''.
These darts encode a set of simplices that contain one another, one for each dimension. Because darts track containment, a set of darts is sufficient to encode the topology of mesh. Navigation through the mesh topology is then performed with sequences of atomic ``switch'' operations that change a single element of these sets.  

To the best of our knowledge, all proposed mesh data structures focus on storing the connectivity and the attributes of a single mesh. One can encode meshes within other meshes just using attributes: the embedded meshes can be stored using tags, for example. However the management of the tags during editing operations is challenging, leading to complex algorithms: an interesting examples are discussed in \cite{FTB:2016:MVR,Thomas2011}, where the link condition check becomes algorithmically complex and expensive.

In our approach we can use any of these data structures to encode individual meshes and we provide a mechanism for keeping these meshes synchronized. We opted for using an extension of the indexed base data structure to store each mesh due to concerns for cache consistency and compactness, but navigate our meshes using the dart abstraction due to its elegance and dimension-agnosticism.

\subsection{Applications}

There is a surprisingly large volume of academic work in graphics and scientific computing using multiple embedded meshes. As we cannot overview all of them due to space considerations, we focus on the applications where heavy changes to multiple mesh topologies are needed and where we believe multimesh could be impactful.

\paragraph{Parametrization}

A discrete mesh parametrization \cite{Floater} is a mapping from a surface to a plane, usually encoded as a pair of triangulations with the same number of triangles but different connectivity to guarantee that the planar triangulation is embeddable on the plane by adding seams. Parametrization algorithms often prescribe seams \cite{Sheffer2006} and then minimize a geometric energy integrated on triangles, for which a face-to-face correspondence is sufficient. Surprisingly few algorithms consider optimizing the seam connectivity directly \cite{LiuFerguson:2017:SSE,Ray2010,Gu2002}.

Multimesh is a natural representation in this setting.
It can independently optimize energies defined on either triangulation by utilizing separate meshes and the topologies are automatically updated after each operation, including the seams (\Cref{sec:applications}).

\paragraph{Periodic Meshing}

Periodic meshes are popular in digital fabrication as a scalable representation of two-scale microstructures \cite{PZMPCZ15,Schumacher2015}. While creating a periodic mesh from scratch is possible by filling the interior of a periodic boundary with a constrained Delaunay method \cite{delaunaybook}, the optimization of a periodic mesh is challenging \cite{cgalperiodic3d,cgalperiodic2d} as the periodicity must be accounted for as operations near periodic boundaries can induces both topological and geometrical changes across these boundaries. 

Multimesh tackles this problem by representing the periodic geometry as both a mesh with a toroidal topology to encode the periodicity and a mesh holding the embedding of a single tile. A 2D square tile is therefore represented as both a square and a topological torus.

\paragraph{Multi-Material Meshing}

The use of explicit surface representations embedded in a volumetric mesh is widely used to simulate scenes with multiple materials. 
Different elements of the mesh are encoded as different materials with different rheologies, such as air, water, or elastic materials.
In fact, the efficacy of such representations for complex geometric flows and fluid simulation have been demonstrated in the seminal series of work on the Deformable Simplicial Complex (DSC) \cite{Misztal2012,misztal2010deformable,Misztal2014,Erleben2011,Misztal2010}.

Although explicitly modeling the entire space makes the simulation code simpler and more accurate, creating and updating the mesh is challenging. 
Substantial efforts have also invested in avoiding to solve this challenge by using other discretizationss by \emph{implicitly} representing the boundaries, such as in particle-based methods like Smoothed-particle hydrodynamics \cite{Monaghan1992,Desbrun1996} or in grid-based methods like the Material-Point method \cite{Sulsky1994,Stomakhin2013}.
These implicit methods, however, come with the cost of not having convergence guarantees and lower accuracy for the same computational budget. A similar approach has been proposed in contact resolution, where an air mesh \cite{Mller2015,Jiang2017} can be used to detect and respond to collisions in lie of traditional collision detection, and in vector graphics, where these meshes are used to define diffusion curves \cite{Orzan2008} and surfaces \cite{Takayama2010}. 

Remeshing meshes with embedded surfaces require a large implementation effort \cite{FTB:2016:MVR} and specialized geometrical and topological conditions \cite{Thomas2011} to ensure the validity of all embedded meshes. To contrast, the multimesh representation explicitly encodes all meshes and automatically keeps them synchronized, encapsulating all the topological difficulties in editing or navigating these meshes.

\section{Formulation}
\label{sec:formulation}

We start by defining high-level concepts used by our data structure. 
A simplex $\s$ is a set of vertices $s$ and its dimension is one less than its cardinality $\dim(\s) = |s|-1$. A simplex $\bar \s$ is a face of $\s$ if $\bar \s \subset \s$ and $\Face(\s)$ is the set of all faces of $\s$.
These faces therefore have lower dimension than $\s$.

\begin{definition}[Mesh]
We define a \emph{mesh} $\K^d$ as a pure, manifold, simplicial complex of dimension $d$. That is, $\K^d$ is a set of simplices that satisfies:
\begin{enumerate}
    \item For every simplex $\s \in \K^d$, $\Face(\s) \subset \K^d$;
    \item For every simplex $\s_1, \s_2\in \K^d$, if $\s_1 \cap \s_2\neq \emptyset$, then $\s_1 \cap \s_2 \in \Face(\s_1)$;
    \item (Pure) For every simplex $\s \in \K^d$, if $\dim(\s)<d$, then there exists a simplex $\s^\prime \in \K^d$ such that $\dim(\s^\prime) = d$ and $\s \in \Face(\s^\prime)$;
    \item (Manifold) Every simplex of dimension $d-1$ is a face of no more than two simplices of dimension $d$.
\end{enumerate}    
\label{def:mesh}
\end{definition}
For convenience, we denote $\mathcal{K}^d$ the set of all meshes of dimension $d$ and denote   a simplex of dimension $d$ in a mesh $\K^d$ as a \emph{facet} (e.g., a triangle in a triangular mesh). 
For further information on simplicial complexes we refer the reader to the book of \citet{Damiand2014}.

\subsection{Mapping Multiple Meshes}
We are now ready to define the \emph{containment map}, which maps simplices from one mesh to another mesh of equal or higher dimension. For instance, in order to have an edge mesh contained in a triangle mesh, we must define a containment map from the simplices of the edge mesh (edges and vertices) to the simplices of the triangle mesh.

\begin{definition}[Containment Map]
Let $\K^k$ and $\L^\ell$ bet two meshes with $k \leq \ell$.
A \emph{containment map} $\Phi_{\K^k}^{\L^\ell}\colon \K^k \rightarrow \L^\ell$ is a map between the simplices of $\K^k$ and $\L^\ell$ that preserves the face relationship. 
That is, for every simplex $\s \in \K^k$, $\Phi_{\K^k}^{\L^\ell}(\Face(\s)) = \Face(\Phi_{\K^k}^{\L^\ell}(\s))$. 

If $\Phi_{\K^k}^{\L^\ell}$ exists then we say that $\L^\ell$ \emph{contains}
$\K^k$ because $\forall \sigma^k \in \K^k, \Phi_{\K^k}^{\L^\ell}(\sigma^\ell) \in \L^\ell$,
which is a partial ordering.
\begin{figure}
\centering
\includegraphics[width=\linewidth]{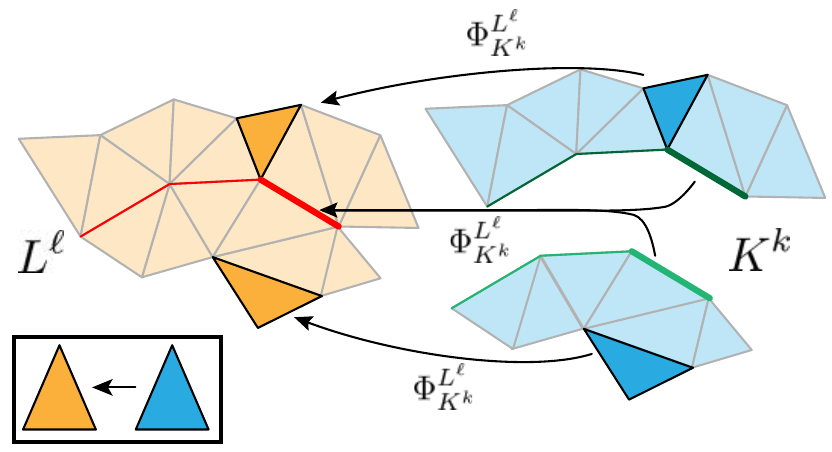}
\caption{Containment map $\Phi_{\K^k}^{\L^\ell}$ from a mesh with seams to a mesh without the seam. Although all of the triangles are mapped uniquely, multiple edges on the boundary can map to the same edge on the root mesh.}
\label{fig:containment}
\end{figure}

\label{def:containment}
\end{definition}
Following the example of the edge mesh in a triangle mesh, not every edge and vertex in the triangle mesh must have a counterpart in the edge mesh (see \Cref{fig:containment}). That is, $\Phi_{\K^k}^{\L^\ell}$ is in general not invertible.
With a collection of meshes and a containment map we can construct a multimesh.

\begin{definition}[Multimesh]
A \emph{multimesh} $\M^d$ is a tree whose nodes are meshes of dimension $\ell \leq d$ and whose edges are containment maps mapping simplices in a child node to its parent. The \emph{root mesh} $\mathcal{R}^d$ is the mesh at the root of this tree.
\end{definition}
\begin{figure}
\centering
\includegraphics[width=\linewidth]{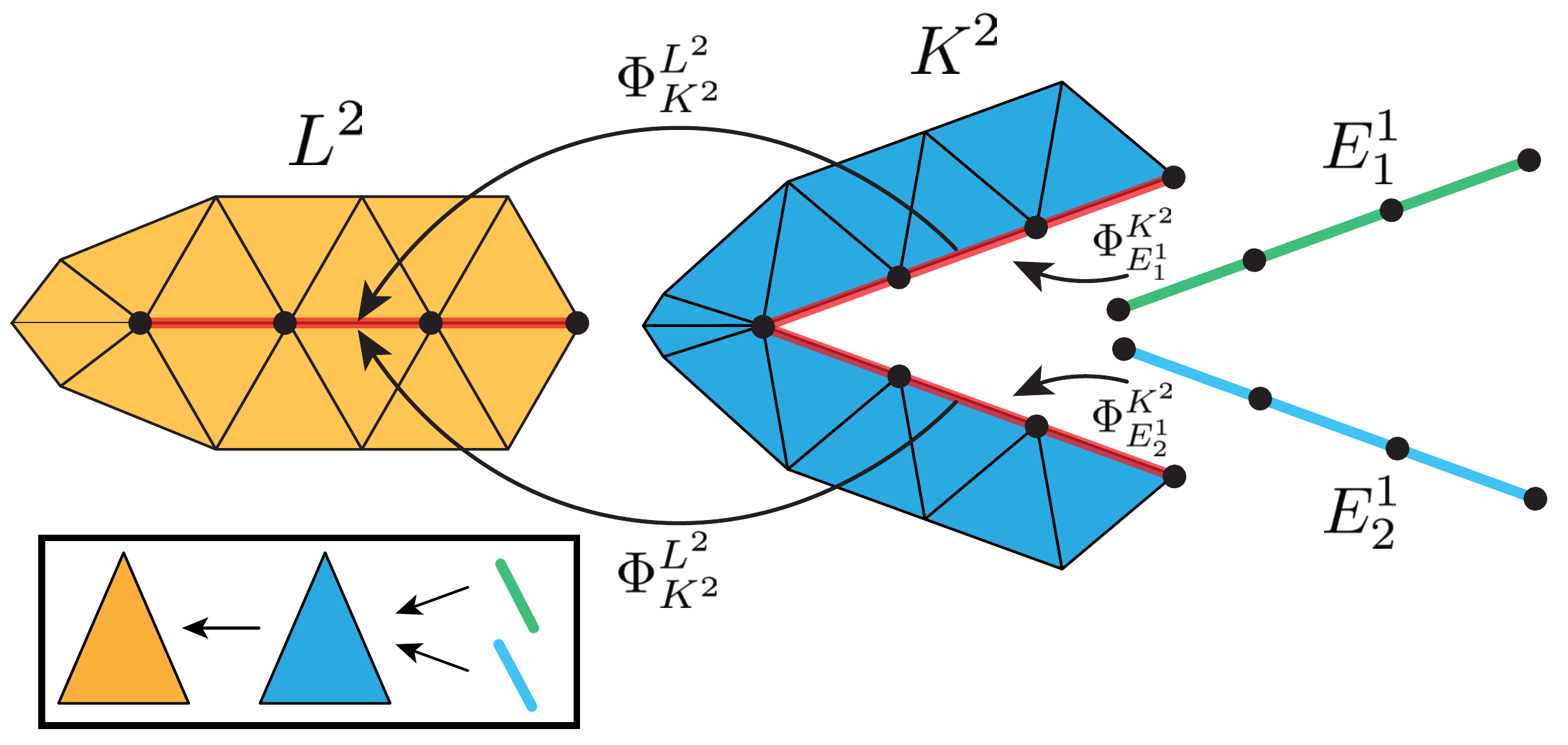}
\caption{A multimesh of depth two, whose leaf nodes are edge meshes $E_1^1$ and $E_2^1$, has a triangle mesh node $K^2$ with a seam (red) and a root mesh $L^2$ without the seam (red).}
\label{fig:multimesh}
\end{figure}

This definition imposes necessary condition for the partial ordering of the multimesh with respect to mesh dimensions: each parent must be of a higher or equal dimension than all its children. 
This, in conjunction with containment maps, induces a tree over simplices where every simplex both has a unique parent simplex in its parent mesh and therefore has a unique root simplex on the root mesh.
This enables efficient predicates for checking whether two simplices are equivalent to one another and for mapping simplices between arbitrary meshes in a multimesh.

\subsection{Topological Operations}
A topological operation transforms a mesh into another mesh by creating or deleting some of its simplices. To ease the notation in this section, we use $\overline{\cdot}$ to denote the mesh after the operation.

\begin{definition}[Topological Operation]
A \emph{topological operation} $\psi^d: \mathcal{K}^d \rightarrow \overline{\mathcal{K}}^d$ of dimension $d$ is a map transforming a mesh $\K^d$ into a mesh $\overline{\K}^d$.
\end{definition}

\begin{figure}
\centering
\includegraphics[width=0.6\linewidth]{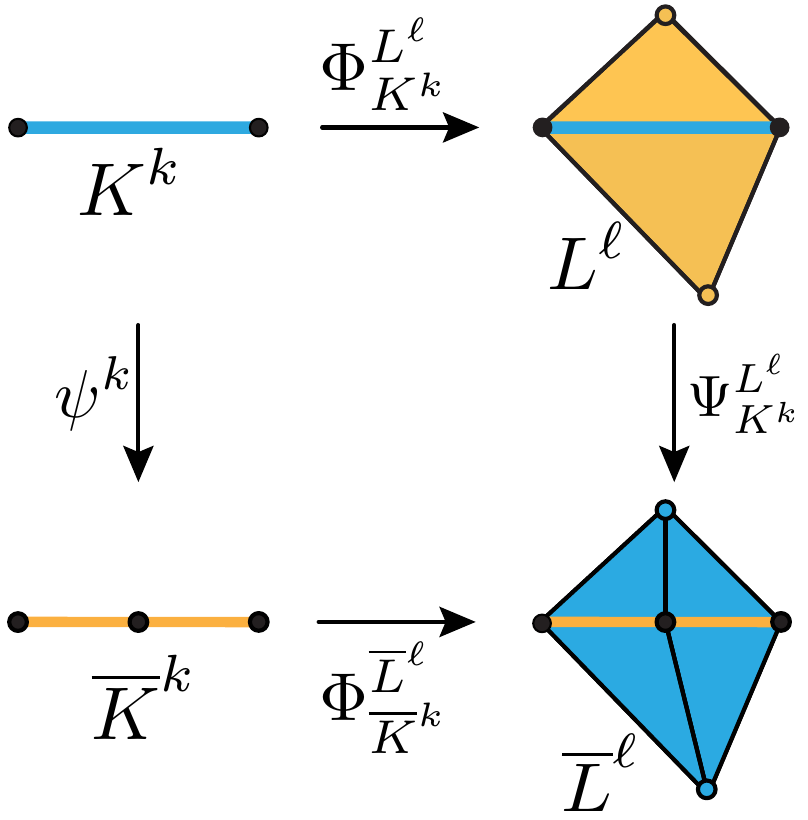}
\caption{The extension $\Psi_{\K^k}^{\L^\ell}$, which is defined in terms of $\psi^k$ defines how the containment map $\Phi_{\overline{\K}^k}^{\overline{\L}^\ell}$ needs to be updated. 
}
\label{fig:operations_update}
\end{figure}

To use topological operations in a multimesh setting, we need to further generalize it so that an operation applied to any node of the tree can be propagated, using the containment maps, to the other nodes. We note that the operation affects nodes above and after in the tree. For example, an edge split on the edge mesh in \Cref{fig:operations_update} will have to be propagated to the triangle mesh linked above it, or a split on a triangle needs to affect all edges below it.

\begin{definition}[Extension]
Let $\psi^k$ be a topological operation transforming a mesh $\K^k$ into a mesh $\overline\K^k$. We denote with $\Psi_{\K^k}^{\L^\ell}$, $k \leq \ell$, the \emph{extension} of $\psi^k$ to a mesh $\L^\ell$ of dimension $\ell$. $\Psi_{\K^k}^{\L^\ell}$ is a topological operation of dimension $\ell$ acting on a mesh $\L^{\ell}$ containing $\K^k$ and which reproduces $\psi^k$. 
That is, if $\L^{\ell}$ contains $\K^k$ and $\overline{\L}^\ell = \Psi_{\K^k}^{\L^\ell}(\K^{k})$ then $\overline \L^\ell$ contains $ \overline\K^k$.
\end{definition}
Note that $\Psi_{\K^k}^{\L^\ell}$ denotes an operation on $\L^\ell$ that extends $\psi^k$ from $\K^k$, and not a function that maps from $\L^\ell$ to $\K^k$.

An example of this extension for an edge split is shown in \Cref{fig:operations_update}, where an operation on an edge is \emph{extended} to a triangle mesh. We note that specific care must be taken to guarantee that a topological operation can be extended on a multimesh.

To define the opposite of an extension, we need to consider the set of child simplices which are mapped from a given simplex $\s$ (\Cref{fig:containment}).

\begin{equation*}
\mathcal{I}^{\K^k}_{\L^\ell}(\s) = \{\s^\prime \in \K^k \mid \Phi_{\K^k}^{\L^\ell}(\s^\prime) = \s\}.
\label{eq:pseudoinverse}
\end{equation*}
We note that this set can be empty: not all simplices in the co-domain of the containment map must correspond to a simplex in the domain. For example, if we have a triangle mesh embedded in a tetrahedral mesh, not all faces of the tetrahedral mesh have a corresponding face in the triangle mesh. More interestingly, $\mathcal{I}^{\K^k}_{\L^\ell}(\s)$ can contain multiple disconnected components: consider a triangle mesh and its corresponding uv-mesh with seams; an edge on the triangle mesh might map to two different edges in the uv-mesh, one for each side of the seam.

\begin{definition}[Restriction]
Let $\psi^\ell$ be a topological operation transforming a mesh $\L^\ell$ into a mesh $\overline\L^\ell$ affecting simplices $\s \in \L^\ell$. We denote with $\Gamma_{\L^\ell}^{\K^k}$, $k\leq l$, the \emph{restriction} of $\psi^\ell$ to a mesh $\K^k$ of dimension $k$ for $\s$ if $\mathcal{I}^{\K^k}_{\L^\ell}(\s) \neq \emptyset$ (if $\mathcal{I}^{\K^k}_{\L^\ell}(\s)$ is empty the restriction does not exist). $\Gamma^{\K^k}_{\L^\ell}$ is a topological operation of dimension $k$ that generates $\overline \K^k$ when $\s$ is mapped to $\K^k$: $\Gamma^{\K^k}_{\L^\ell}(\mathcal{I}^{\K^k}_{\L^\ell}(\s)) = \overline\K^k$.
\end{definition}

\paragraph{Topological Operations and Containment Maps.} 

We now extend the effect of topological operations to containment maps (\Cref{fig:operations_update}).

\begin{definition}[Containment Map Update]
Let $\K^k$ and $\L^{\ell}$, $k\leq \ell$, be two meshes, $\Phi^{\L^\ell}_{\K^{k}}$ a containment map, and $\psi^k$ a topological operation with extension $\Psi_{\K^k}^{\L^\ell}$.
Similarly to Definition \ref{def:containment}, we define the \emph{Containment Map Update} as the identity map between $\overline{\K}^k= \psi^k(\K^k)$ and the corresponding subset of the modified mesh $\overline{\L}^\ell = \Psi_{\K^k}^{\L^\ell}(\Phi^{\L^\ell}_{\K^{k}}(K^k))$, which we denote with a convenient overload of notation as $\Phi_{\overline\K^k}^{\overline\L^\ell}$. 
\end{definition}

\paragraph{Topological Operations on a Multimesh}

Equipped with these definitions, we can now define the effect of a topological operation on a multimesh. A local operation applied to a mesh $\K^k$ within a multimesh $\M^d$ has a global effect on the entire multimesh and can even lead to a non-trivial behaviour on $\K^k$ itself. Consider the example in \Cref{fig:multimesh}: an edge of $E^1$ is split, this triggers a change in topology in all other meshes.

The algorithm for applying a topological operation $\psi^k$ to a mesh $\K^k$ in a multimesh is composed of 3 stages:
\begin{enumerate}
    \item Collect the affected simplices $\s$ and use the containment function $\Phi_{\K^k}^{\mathcal{R}^d}$ (apply containment maps to the root of the tree) to compute $\mathcal{I}^{\K^k}_{\mathcal{R}^d}(\s)$ (i.e., the corresponding affected region in $\mathcal{R}^d$) and its extension $\Psi_{\K^k}^{\mathcal{R}^d}$.
    \item For each mesh $\L^\ell$ of the tree apply the restriction operation $\Gamma_{\mathcal{R}^d}^{\L^\ell}$ to $\L^\ell$.
    \item For all edges in the tree, update the corresponding containment map.
\end{enumerate}

\paragraph{Constraints}

Thus far we have described how to develop and define operations that are compatible between different meshes, but not every operation can be executed on every mesh. This can be due to topological constraints on an individual mesh (link condition~\citep{PACHNER1991129}) or on the multimesh validity (see collapse operation in Section \ref{sec:supported_ops}). To perform an operation, we thus first ensure that individual operations on the meshes can be executed (at minimum, the link condition needs to pass on all meshes) and then the operation can be executed on the multimesh. 
\section{Implementation}

The multimesh approach can be implemented on top of any existing single mesh data structure, as long as it supports the encoding of meshes of different dimensions. We describe here the general approach for implementing, assuming the usual features available in a mesh data structure: storing of simplices and their attributes, plus a way to navigate on each mesh.

\subsection{Encoding}

\begin{figure}
\includegraphics[width=.6\linewidth]{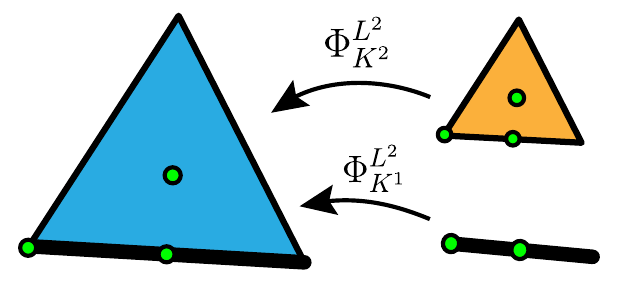}
\caption{Containment map from a $2$-dart and $1$-dart to the same $2$-dart.}
\label{fig:containment_map_1}
\end{figure}

In our implementation, we rely on the notion of \emph{dart}~\citep{Edmonds1960} for navigation and encoding the mapping. For a simplex $\s^k\in {\K^k}$ we define the dart as a collection of simplices ``lower'' than $\s^k$; that is, $\xi(\s^k) = \{\s_m\}_{m=0}^k$ where $\s_m$ is a $m$-simplex and $\s_n \subset \s_m$ for any $n < m$. To ease the nation, where not ambiguous, we use $\xi$ instead of $\xi(\s^k)$.
Rather than store $\Phi_{\K^k}^{\L^\ell}$ for each simplex $\s \in \K^k$ we store two darts  for each facet of $\K^k$, we represent the map
\[
\bar \xi_j = \{ \Phi_{\K^k}^{\L^\ell}(\xi_j) \}
\]
for $j < k$, where $\bar \xi_j$ and $\xi_j$ are the $j$-th simplex in $\bar \xi$ and $\xi$ respectively (\Cref{fig:containment_map_1}).

\begin{figure}
\includegraphics[width=.6\linewidth]{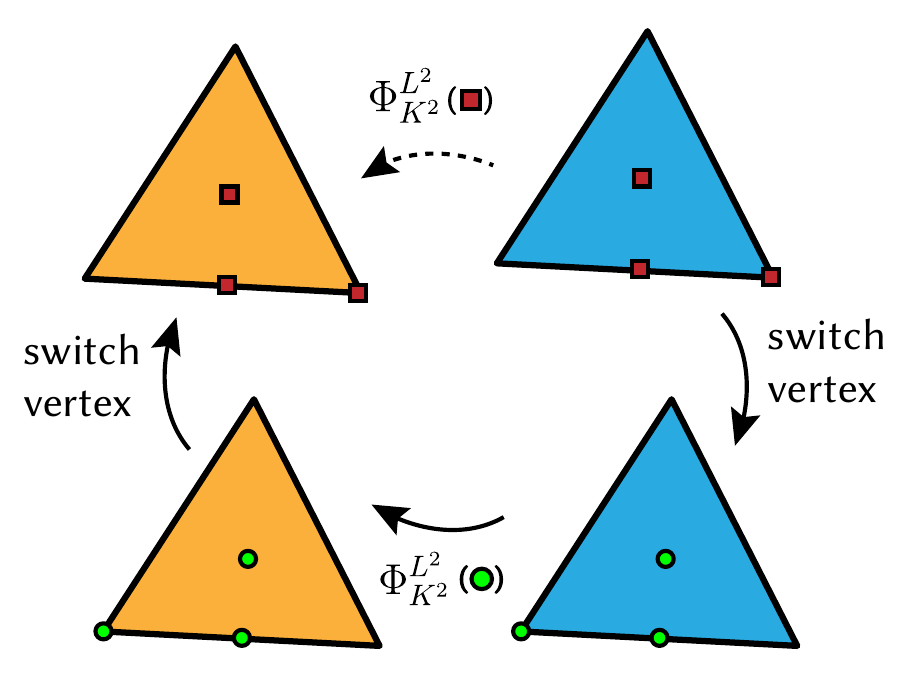}
\caption{In a single simplex the containment map of an arbitrary dart (red square) is navigating to an anchor (green square) with switches (here switching a vertex), reading the other dart in the anchor, and then inverting the navigation.
}
\label{fig:anchor_switch_commutative}
\end{figure}

This is stored as an attribute on both $\s$ and on all simplices in the parent mesh 
$\Phi_{K^k}^{L^\ell}(\s)$.
This allows us to efficiently map a dart (and a simplex) between $\L^\ell$ and $\K^k$ for each facet of $\K^k$. The dart formalism \cite{Damiand2014} comes with a canonical means of navigating between faces of a single simplex (called \emph{switch} \cite{Brisson1989}: with this local navigation, is straightforward to extend the mapping from a single dart in a simplex to all its other darts (\Cref{fig:anchor_switch_commutative}). By extending this construction on the entirety of $\K^k$, this fully encodes 
$\Phi_{\K^k}^{\L^\ell}$.

\subsection{Construction}
\label{sec:construction}
We construct our maps in two ways: we can either pass the containment map for every vertex in each simplex directly or we can construct them from tags.

\begin{figure}
\includegraphics[width=.6\linewidth]{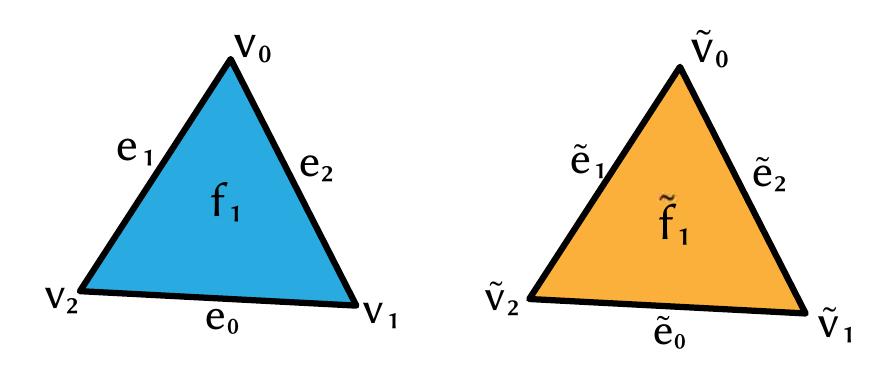}
\caption{
Because we know 
$v_0 \leftrightarrow \tilde v_0$,
$v_1 \leftrightarrow \tilde v_1$,
$v_2 \leftrightarrow \tilde v_2$ we also know 
$e_0 \leftrightarrow \tilde e_0$,
$f_1 \leftrightarrow \tilde f_1$ and so the dart $(\{\tilde v_1, \tilde e_0, \tilde f_1\},(\{v_1,e_0,f_1\})$ is a valid anchor.
}
\label{fig:obj_mesh}
\end{figure}

\paragraph{Construction from facet bijection}
In this method we obtain two meshes with the same number of facets and each facet in one mesh is paired with one in the other.
Furthermore, for each of these paired facets we know which vertices correspond to one another.
This correspondence between vertices in a single simplex defines a correspondence between every face of that simplex, and therefore one can construct an anchor between these two facets (\Cref{fig:obj_mesh}).

\begin{figure}
\includegraphics[width=.6\linewidth]{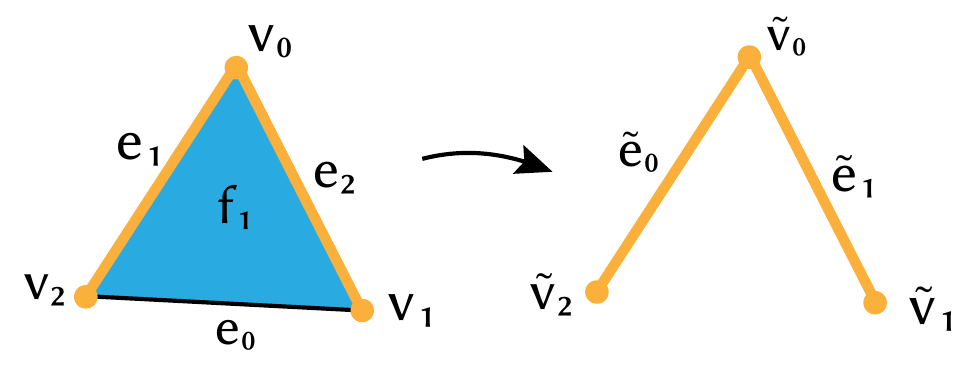}
\caption{
If the top two edges of a triangle are tagged we know that 
$v_0 \leftrightarrow \tilde v_0$,
$v_1 \leftrightarrow \tilde v_1$,
$v_2 \leftrightarrow \tilde v_2$ and
$e_1 \leftrightarrow \tilde e_0$,
$e_2 \leftrightarrow \tilde e_1$; therefore 
$(\{\tilde v_2, \tilde e_0\},\{v_2,e_1,f_1\})$ and 
$(\{\tilde v_0, \tilde e_1\},\{v_0,e_2,f_1\})$ are valid anchors.
}
\label{fig:tag_mesh}
\end{figure}

\paragraph{Construction from tags}
Given a tag on $\L^\ell$, for each tagged $k$-simplex $\s^{k} \in \L^\ell$ ($k \leq \ell$), we construct a mesh $\K^k$ by collecting all $\s^{k}$.
As we construct $\K^k$, we record the vertices of $\s^k$ and keep track of how every vertex in $\s^{\ell}$ maps to the vertices of $\s^k$. We use this information to construct an anchor 
$(\bar \xi_j,\xi_j)$ (\Cref{fig:tag_mesh}).

\subsection{Updating the containment map}
\begin{figure}
\includegraphics[width=.6\linewidth]{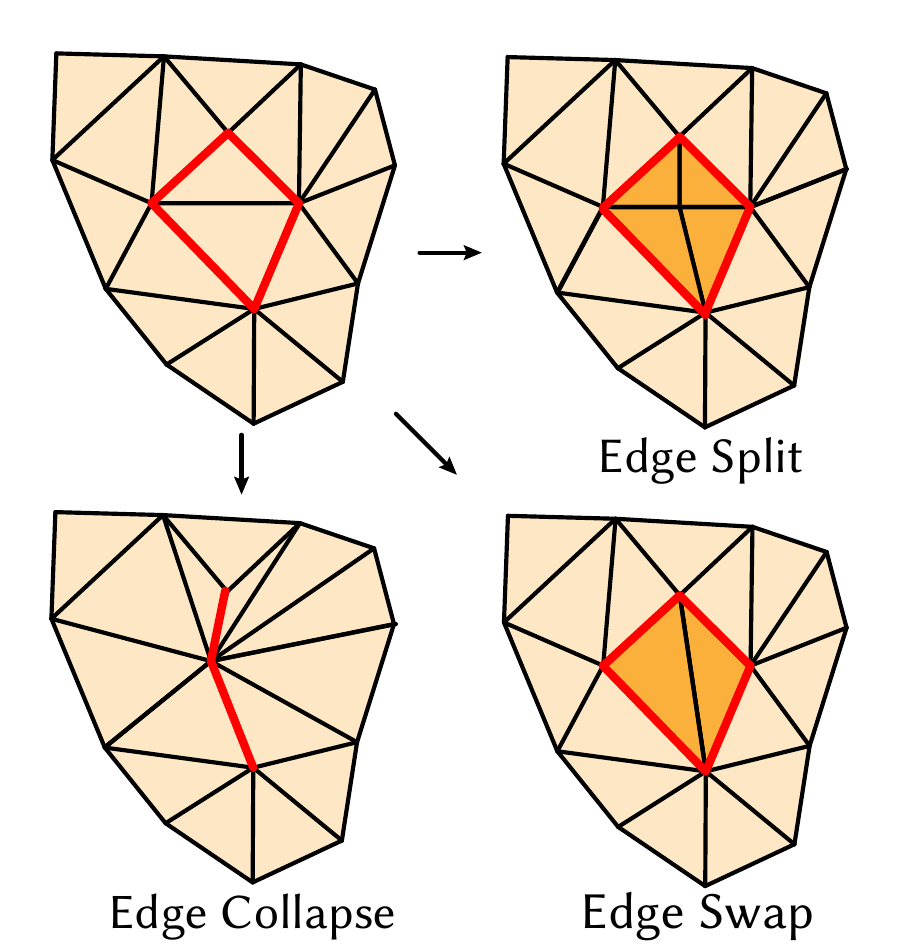}
\caption{
The operations we will discuss, with the boundary of each operation in red.
}
\label{fig:operations}
\end{figure}

Each topological operation requires its own custom mechanism for updating the containment map.
For our particular reference implementation we utilized edge splits, edge collapses, and swaps (\Cref{fig:operations}). Before we discuss mapping changes on the interior of an operation, we need to discuss updates to the boundary of an operation.

\begin{figure}
\includegraphics[width=.6\linewidth]{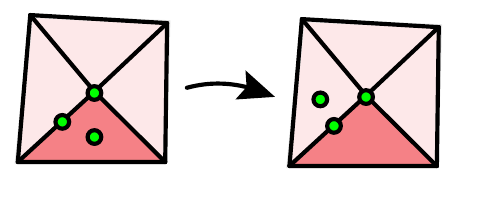}
\caption{
If an anchor on a boundary edge of an operation involves a facet inside of an operation (red) then we find an alternate facet that shares its boundary.
}
\label{fig:infringing_dart}
\end{figure}

\paragraph{Generic Boundary Update Rule}
Every operation has a boundary, and there can be cases where the containment of some facet $\s$ on $\K^k$ can lie on the boundary of an operation on $\L^\ell$. When a simplex is deleted, a simplex in the anchor for mapping $\s$ might also disappear.
If a face of $\xi$ is on the operation boundary, then we change the dart $\xi$ used to store the anchor. If the old anchor was $(\xi, \bar \xi)$, then the new anchor $(\tilde \xi, \bar \xi)$ must, for every $j \leq k$, have the same $j$-simplex for $\tilde \xi$ as $\xi$ (\Cref{fig:infringing_dart}).  
As long as our operations maintain the pureness property of our meshes there will always be a facet available for building an anchor.

There is a special case for this update on the boundary of the mesh $\L^\ell$: if the containment of $\s \in \K^k$, $\bar \s = \Phi_{\K^k}^{\L^\ell}(\s)$ lies on the boundary of $\L^\ell$ there might not be a facet that both includes $\bar \s$ and lies outside of the operation before the operation takes place. 
Each topological operation must account for this special case in an ad-hoc basis to ensure the containment map is correctly updated.

\subsection{Supported Operations}
\label{sec:supported_ops}
We will discuss how to implement the containment map update for split, collapse, and swap.
Note that this construction can be extended to any arbitrary local operations.

\paragraph{Operation 1: Edge Split}
\begin{figure}
\includegraphics[width=\linewidth]{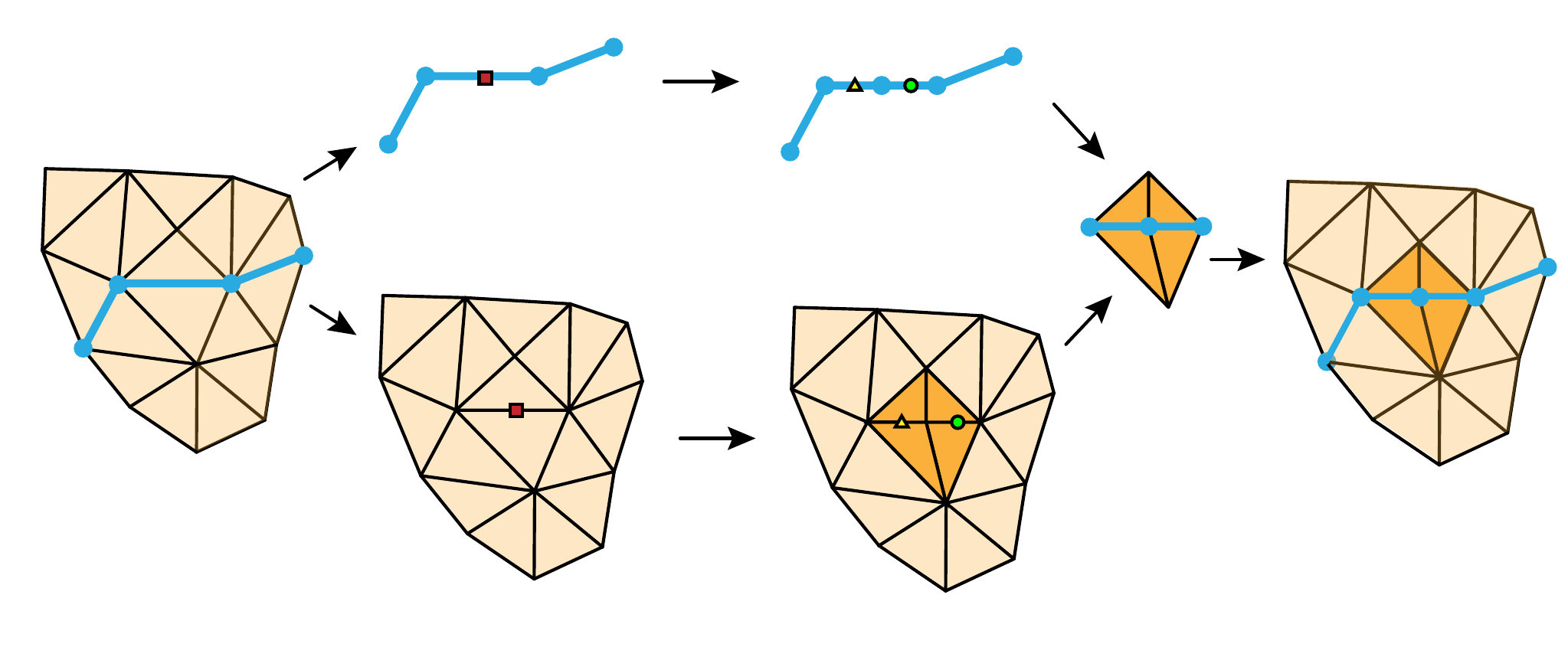}
\caption{
A split transforms splits a mapped edge (red square) into two simplices which can be identified by their endpoints (yellow triangle and green circle). By tracking the endpoints we can reconstruct the correspondence between simplices.
}
\label{fig:split}
\end{figure}
When we split an edge $e$, we substitute every facet for whom $e$ is a face of to two facets, including $e$ itself (e.g., for a triangle mesh, the edge $e$ and the two adjacent triangles, \Cref{fig:split}).
Furthermore, if a simplex $\s$ where $e \in \Face(\s)$ has nonempty $\{\bar \s_j\} = \mathcal{I}_{\L^\ell}^{\K^k}(\s)$, then the $\{\bar e_m\} =\mathcal{I}_{\L^\ell}^{\K^k}(e)$ of the edge also exists. Because all edges $\bar e_m$ will be split and each $\bar \s_j$ will have one $\bar e_m$ as a face, $\bar \s_j$ will be split as well.
Let $a,b$ be the two vertex endpoints of $e$, $c$ the new vertex, and $\s_a$ and $\s_b$  be the two simplices generated by splitting $\s$.
These two simplices can be defined as the sets that replace either $a$ or $b$ endpoint with the new vertex $c$: 
\begin{equation*}
\s_b = \s \setminus \{ a \} \cup \{ c \},
\end{equation*}
and $\s_a$ naturally follows as well.
Note that because $c$ is the only new vertex in the configuration, we know how to map every vertex but $c$.
Let $\bar \s$ be the simplex such that $\Phi_{\K^k}^{\L^\ell}(\bar \s) = \s$, 
and note that split also introduces a single new vertex $\bar c$, and that will be mapped to $c$.
As such, we know $\Phi_{\K^k}^{\L^\ell}$ for every vertex of the simplices $\s_a,\s_b$ to $\bar \s_a,\bar \s_b$ respectively so we can generate an arbitrary dart by taking pairs of sets of corresponding vertices to generate our new anchor.
To update the boundary, because an edge split always replaces each facet adjacent to the boundary of the operation with a new facet, we can update the anchors of a mapped face of $\s$ with an anchor using either $\s_a$ or $\s_b$.

\paragraph{Operation 2: Edge Collapse}
\begin{figure}
\includegraphics[width=\linewidth]{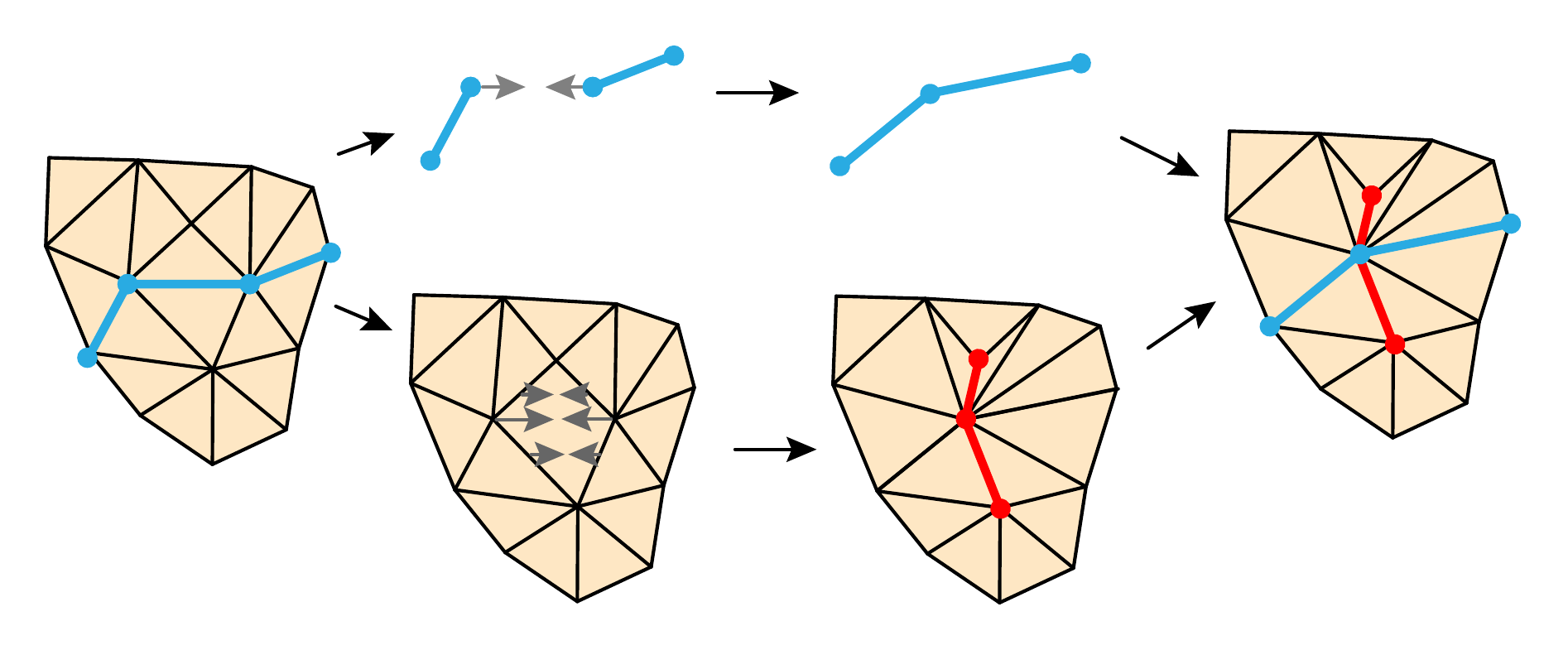}
\caption{
An edge collapse merges the two endpoints of an edge together, also merging the facets that use one of those endpoints (red edges).
}
\label{fig:collapse}
\end{figure}

An edge collapse of edge $e$ removes every facet that has  $e$ as a face and merges the endpoints of $e$ into a single vertex.
Once again, let  $a,b$ the endpoints of $e$ and let $\s$ be a facet for which $e \in \Face(\s)$ and note that conceptually the collapse is the merging of $a$ and $b$ into a single vertex.
In the case of a collapse, every $\s$ is removed, so the only simplices that remain are on the boundary of the operation, like the red edges in \Cref{fig:collapse}.

The collapse operation deletes several facets without replacing them, and therefore has an inherent risk of breaking the pureness of a mesh.
We filter invalid operations with the link condition to guarantee that the mesh after the operation after the mesh will remain pure, hence guaranteeing new facets for updating anchors.
The only change to our generic update rule is that for each merged simplex we must look at the neighbors of both of the simplices that were merged.

\paragraph{Operation 3: Swaps}
Swaps are implemented as sequences of splits and collapses \cite{Diazzi2023} and, thus, do not require any special update rules.

\section{Results}
\label{sec:applications}

\begin{figure}
    \centering
    \includegraphics[width=\linewidth]{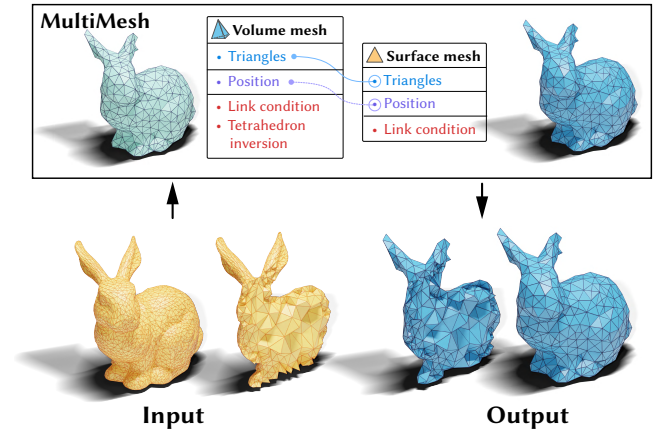}
  \caption{A multimesh composed of a volume (tetrahedra) and its surface (triangle). Isotropic remeshing~\cite{Botsch2004} is applied to the triangle mesh, and the operations are automatically propagated to the volumetric mesh. The tetrahedra on the interior are mostly unchanged because we only perform operations on the surfaces.}
    \label{fig:laplacian}
\end{figure}

We implemented the multimesh using the open-source mesh library Wildmeshing toolkit~\cite{toolkit}. 
We implemented four different mesh optimization algorithms to illustrate the power and simplicity of multimesh. For every example, we explain which meshes are involved and use a diagram to illustrate the multimesh connection. We use blue edges to depict the containment map (a solid circle illustrates the parent mesh), purple for attributes in each mesh, with purple edges to specify how attributes are updated, and red for the invariants of each mesh. 
For instance, the diagram in \Cref{fig:laplacian} shows a multimesh connecting some triangular faces of the tetrahedral mesh with the surface (blue edge), explains that the position of the child mesh (surface) is updated from the parent (tetrahedral mesh), and both meshes have some invariants.

\subsection{Validation}
To validate our implementation of the multimesh, we re-implemented TetWild~\citep{Hu:2018:TMW:3197517.3201353}. 
The algorithm constructs a conforming tetrahedral mesh for an input triangle mesh by: (1) inserting triangles within a background mesh within a bounding box, (2) optimizing the quality of the tetrahedral mesh while preventing the inserted surface and open boundaries from moving outside of an envelope~\citep{Wang2020}.
The original algorithm uses several tags to keep track of surfaces embedded in the mesh and their boundary, and preserving the tags during the local operation is one of the major challenges in implementing the algorithm. With multimesh, we construct a mesh for every structure we want to preserve. That is, the tetrahedral mesh is the root mesh that has three children: the boundary (bounding box) triangle mesh, the surface of the inserted triangle mesh, and the open boundary edge mesh of the inserted surface (Figure \ref{fig:teaser}). With this hierarchy, we implement TetWild by preserving all the child meshes with an envelope. This does not require special handling of the open boundary (edges) or of the boundary (triangles). Every mesh has its dedicated envelope and projection and is \emph{automatically} updated by operations acting on the main tetrahedral mesh. 

\begin{figure}
    \centering
    \includegraphics[width=\linewidth]{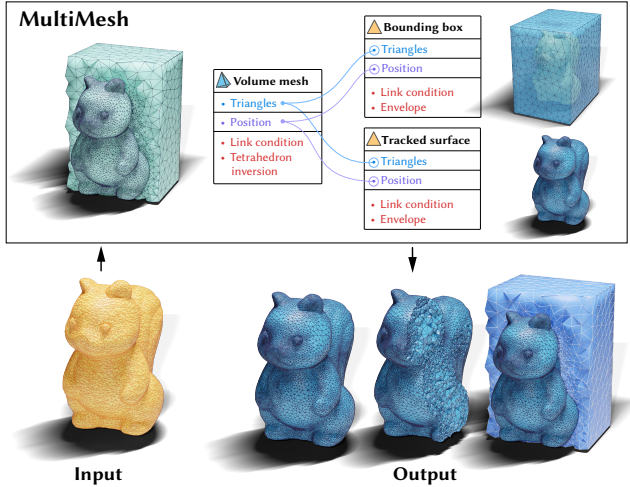}
  \caption{A multimesh composed of a volume (tetrahedra), the embedded surface (triangles), and its bounding box (triangles). TetWild~\cite{Hu:2018:TMW:3197517.3201353} is applied to the tetrahedral mesh, and the operations are automatically propagated to all other meshes that are used for envelope containment.}
  
    \label{fig:tetwild}
\end{figure}

\paragraph{Topology Preservation.}
Additionally, the multimesh allows the preservation of the tracked surface and the open boundary's topologies, a feature not available in the original TetWild algorithm \cite{Hu:2018:TMW:3197517.3201353}.  If the input is water-tight only two triangle sub-meshes are needed to initialize an envelope, one for the bounding box and one for the tracked surface (\Cref{fig:tetwild}). 
If the mesh contains an open boundary (\Cref{fig:teaser}), we construct an edge sub-mesh as well to initialize an edge envelope.

\begin{figure}
    \centering\footnotesize
    \parbox{.02\linewidth}{\centering\rotatebox{90}{Time (s)}}
    \parbox{.47\linewidth}{\includegraphics[width=\linewidth]{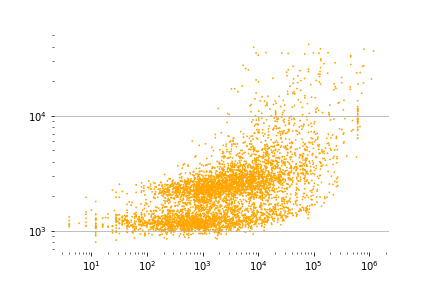}}\hfill\hfill
    \parbox{.02\linewidth}{\centering\rotatebox{90}{Number of models}}
    \parbox{.47\linewidth}{\includegraphics[width=\linewidth]{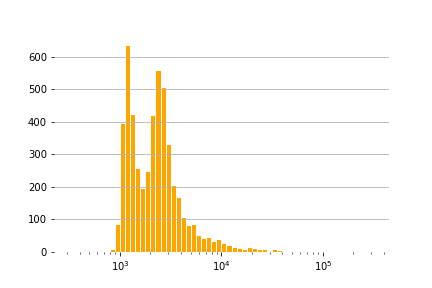}}
    \parbox{.02\linewidth}{~}
    \parbox{.47\linewidth}{\centering Number of input faces}\hfill\hfill
    \parbox{.02\linewidth}{~}
    \parbox{.47\linewidth}{\centering Time (s)}
    \caption{Statistics of the run time of our re-implementation of TetWild on the 5097 manifold models form the Thingi10k dataset.}
    \label{fig:large-scale}
\end{figure}

\paragraph{Large Scale Dataset}
We run the experiments on a Xeon E5-2690 v2 @ 3.00GHz with a maximal time limit of 12 hours. \Cref{fig:large-scale} shows the statistics of running our implementation on the manifold subset of the Thingi10k dataset containing $5163$ models. 
Though the algorithm is slower than TetWild, recall that it preserves the topology of the input surfaces. It produces volumetric meshes with maximal AMIPS distortion below 150 for 5097 models, and the remaining 66 do not terminate within 12 hours.

\subsection{Applications}

\begin{figure}
    \centering
    \includegraphics[width=\linewidth]{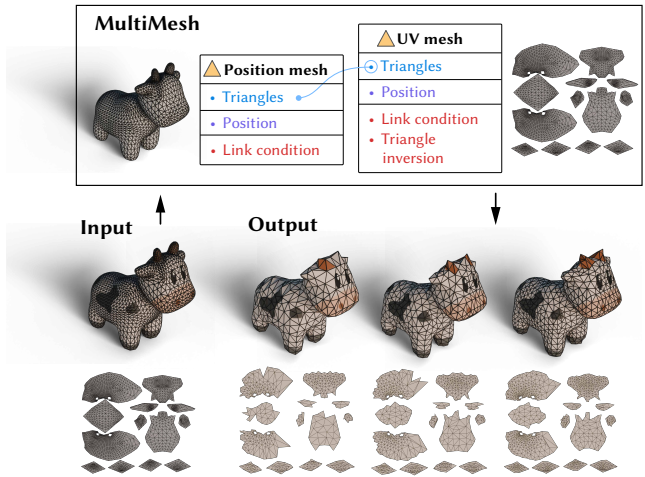}
    \caption{Example of multimesh for optimizing a parameterization. The multimesh is comprised of a seamed mesh to encode texture coordinates (UVs) and a seam-free mesh to encode positions. We run shortest edge collapse~\citep{Hoppe1996} to coarsen the shape.}
    \label{fig:parametrization}
\end{figure}
\paragraph{Surface mesh optimization with seams.} This is a classical example in texture mapping: we have a triangular mesh and its corresponding uv mesh. The triangular mesh is the root of the multimesh that has only one child, the uv mesh. Each triangle comes with ordered indices for the position mesh and the uv mesh, so we use our facet bijection method for creating simplices~\Cref{sec:construction}.
\Cref{fig:parametrization} shows an example of such optimization. The setup is straightforward: we use standard shortest edge collapse~\citep{Hoppe1996} on the parent mesh and add an invariant to prevent triangles from inverting on the child mesh. The multimesh structure automatically carries the operations on both meshes while maintaining a correspondence between the faces and preserving the seam structure in the uv mesh.

\begin{figure}
    \centering
     \includegraphics[width=\linewidth]{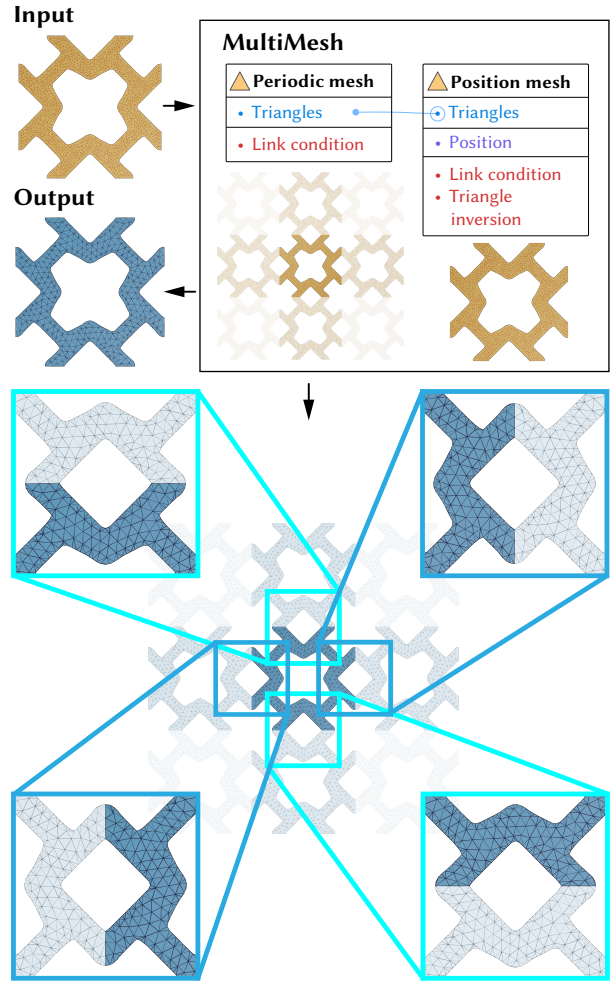}
    \caption{Example of multimesh for periodic meshing. A multimesh stores two meshes: one with the topology of a disk and one with the topology of a torus. We optimize the geometry of the individual tile, and multimesh automatically keeps the periodic boundaries consistent as it updates the connectivity of the torus.}
    \label{fig:periodic}
\end{figure}

\begin{figure}
    \centering
    \includegraphics[width=\linewidth]{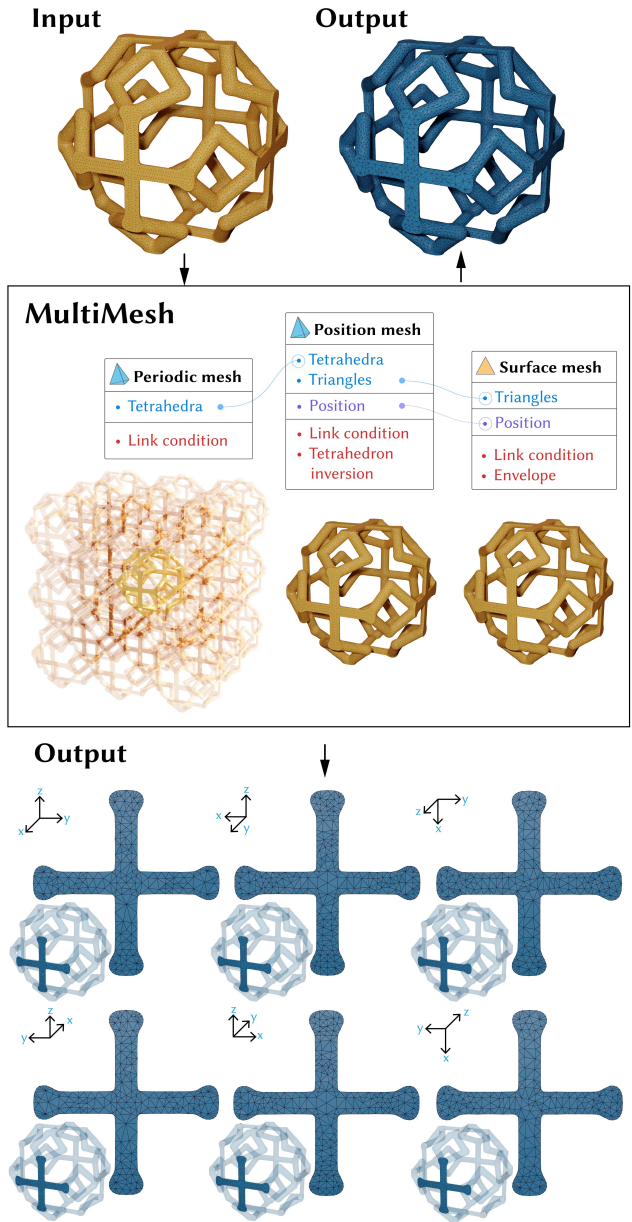}
   \caption{Example of multimesh for 3D periodic meshing. A multimesh stores two tetrahedral meshes and one triangle mesh: one tetrahedral mesh with the topology of a micro-structure cell, one tetrahedral mesh with the periodic topology, and a triangle mesh track the surface of the micro-structure that stays in an envelope during the optimization. We optimize the geometry of the individual micro-structure, and multimesh automatically keeps the periodic boundaries consistent. The three pairs of crosses show the consistent periodic boundaries after the optimization.}
    \label{fig:periodic3d}
\end{figure}

\paragraph{Periodic mesh optimization.} We optimize the quality of a triangular and tetrahedral periodic tiles (\Cref{fig:periodic} and \Cref{fig:periodic3d}). 
The challenge is to maintain the tileability of the mesh as we perform operations to improve its quality or coarsen the mesh while maintaining good quality. With the multimesh,  by merging the corresponding periodic simplices in connectivity we construct a periodic topology of the input tile as a root mesh and use the input tile mesh as a child. 
The periodic mesh has no embedding and primarily guarantees that operations on both sides of a period are synchronized. As in the uv-parameterization example, the mapping across meshes is specified by the facet bijection method for construction~\Cref{sec:construction}. To improve the quality of the tile, we perform isotropic remeshing on the child mesh in two dimensions and the optimization process of TetWild in three dimensions.
We exploit the extension and restriction of the operation to \emph{automatically} maintain the same geometry across the periodic boundaries.

\paragraph{Surface optimization.} For this example, we aim to optimize the quality of a surface embedded in a tetrahedral mesh while maintaining the validity of the tetrahedral mesh.
In particular, we guarantee that every tetrahedron maintains positive volume. 
We construct the multimesh by first extracting the boundary of the tetrahedral mesh and then assigning the tetrahedral mesh as the root and the boundary surface as the child. 
In this case, the containment map from the boundary surface to the tetrahedral mesh is provided by a tag indicated which simplices lie on the boundary of the tetrahedral mesh. To improve the quality of the surface (\Cref{fig:laplacian}), we perform isotropic remeshing with Laplacian smoothing on the surface while the multimesh carries the updates to the tetrahedra. To ensure the validity of the tetrahedra, we add an  invariant that identifies element inversions after every operation  so we can roll back any topological updates if an inversion occurs.

\subsection{Comparison}
A natural alternative for the multimesh consists of using tags to keep track of any sub-mesh. This idea seems effective and deceptively straightforward to implement: however, this is not the case.

We consider the algorithm proposed in \cite{Vivodtzev2010} as an example. The paper proposes a popular multi-material link condition, which is used in the remeshing package in CGAL \cite{FTB:2016:MVR}.
The paper proposes a link condition that accounts for all the substructures embedded in a multi-material mesh by connecting the tagged substructures to a vertex at infinity and extending the traditional single mesh link condition to this more involved case.
The pseudo-code in \cite[Algorithm 1]{Vivodtzev2010} is complex (41 lines of pseudo-code) and additionally requires the implementation of non-trivial navigation operations on each of the tagged substructures. The algorithm is simple to implement in a multimesh and reduces to collecting all the instances of a given simplex on all meshes in a multimesh and recursively calling individual link conditions in each (\Cref{alg:link}).

\begin{algorithm}
\caption{Pseudo code for multimesh link condition, for the root mesh $\mathcal{R}^d$, a mesh $\K^k$, and an edge $e \in \K^k$.}
\label{alg:link}
\begin{algorithmic}
\Procedure{LinkConditionMM}{$\mathcal{R}^d, \K^k, e$}
\State $e_{\mathcal{R}} \gets \Phi_{\K^k}^{\mathcal{R}^d}(e)$ \Comment{Map $e$ to the root mesh}
\State \Return {\Call{LCInternal}{$\mathcal{R}^d, e_{\mathcal{R}}$}}
\EndProcedure
\State

\Procedure{LCInternal}{$\L^\ell, e$}
\If{not \Call{LinkCondition}{$\L^\ell, e$}}
\State \Return \texttt{False}
\EndIf
\For {$C^c \in $ children of  $\L^\ell$}
    \For {$e_c \in \mathcal{I}_{\L^\ell}^{C^c}(e)$} \Comment{Map $e$ to child mesh}
        \If{not \Call{LCInternal}{$C^c, e_c$}}
        \State \Return \texttt{False}
    \EndIf
    \EndFor
\EndFor
\State \Return \texttt{True}
\EndProcedure
\end{algorithmic}
\end{algorithm}

\section{Concluding Remarks}

We have introduced a novel construction for representing multiple meshes of different dimensions related by containment maps.
Furthermore, we have shown how it can be effectively applied to a variety of meshing applications.

The main drawback of our method is the high implementation complexity and the requirement to have a generic data structure supporting navigation and editing of meshes of multiple dimensions. To simplify reproducibility, we will release a reusable, reference implementation of our data structure and of the applications demonstrated in the paper.

Finally, we believe this approach is more natural than tagging substructures inside meshes or keeping ad-hoc relationships between them. Similar to how a mesh data structure abstracts away tedious details for mesh algorithms and therefore allow for algorithms to be implemented more compactly and elegantly, our data structure will have a similar effect for many geometry processing algorithms that work on more than a mesh at once.

\bibstyle{ACM-Reference-Format}
\bibliographystyle{ACM-Reference-Format}
\bibliography{00-multimesh.bib}


\end{document}